\documentclass[graybox]{svmult}

\usepackage{mathptmx}       
\usepackage{helvet}         
\usepackage{courier}        
\usepackage{type1cm}        
\usepackage{makeidx}         
\usepackage{graphicx}        
\usepackage{multicol}        
\usepackage[bottom]{footmisc}
\usepackage{rotating}

\makeindex             

\begin{document}

\title*{Mobile-Based Experience Sampling for Behaviour Research}
\author{Veljko Pejovic, Neal Lathia, Cecilia Mascolo and Mirco Musolesi}
\institute{Veljko Pejovic \at Faculty of Computer and Information Science, University of Ljubljana, Slovenia \\\email{Veljko.Pejovic@fri.uni-lj.si}
\and Neal Lathia \at Computer Laboratory, University of Cambridge, United Kingdom \\ \email{neal.lathia@cl.cam.ac.uk}
\and Cecilia Mascolo \at Computer Laboratory, University of Cambridge, United Kingdom \\\email{cecilia.mascolo@cl.cam.ac.uk}
\and Mirco Musolesi \at Department of Geography, University College London, United Kingdom \\\email{mirco.musolesi@ucl.ac.uk}
}
%
%
\maketitle

\abstract{The Experience Sampling Method (ESM) introduces in-situ sampling of human behaviour, and provides researchers and behavioural therapists with ecologically valid and timely assessments of a person's psychological state. This, in turn, opens up new opportunities for understanding behaviour at a scale and granularity that was not possible just a few years ago. The practical applications are many, such as the delivery of personalised and agile behaviour interventions. Mobile computing devices represent a revolutionary platform for improving ESM. They are an inseparable part of our daily lives, context-aware, and can interact with people at suitable moments. Furthermore, these devices are equipped with sensors, and can thus take part of the reporting burden off the participant, and collect data automatically. The goal of this survey is to discuss recent advancements in using mobile technologies for ESM (mESM), and present our vision of the future of mobile experience sampling.}

\section{Introduction}
\label{sec:introduction}

Human behaviour often depends on the context in which a person is. This context is described by our physical environment, for example, a semantic location, such as home or work, our physical state, such as running or sleeping, but also by our internal state, for example, our cognitive load. The manifestations of human behaviour are complex, and can be observed through our actions, thoughts, and emotions, to name a few descriptors. For psychologists, understanding human behaviour necessitates capturing behaviour as it happens. Initial methods of capturing behaviour included lab studies, where participants were placed in an artificial situation and closely monitored, as well as retrospective interview studies, where participants were asked to recall their past experiences. However, since behaviour depends on the context, which is often much richer than anything that can be created in the lab, these studies cannot be used to faithfully replicate natural behaviour. \textit{The Experience Sampling Method (ESM)}\index{Experience sampling method (ESM)}\index{ESM} was developed to capture human behaviour as it happens~\cite{Hektner2006}.

The essence of ESM lies in occasional querying of users who then provide immediate answers to questions asked. The method avoids both direct interaction with a researcher/therapist, as well as artificial lab-made environments. As such, ESM-obtained data are, first of all, recorded in the context, and thus of higher ecological validity than data obtained by legacy means of collection. Second, they are recorded closely after the moment of querying, minimising the retrospective bias symptomatic to data harvested by earlier methods. Furthermore, ESM allows long-term querying and longitudinal studying of participants, and may be able to capture samples during infrequently occurring events.

The original approach to sampling users in an ESM study included a programmable beeper\index{Beeper sampling} that indicates times at which a sample should be taken, and a paper diary that participants fill out once the beeper rings~\cite{Hektner2006}. Different forms of data collection and querying, such as calling users on their mobile phones, or using a personal digital assistant (PDA) device to collect data, have been used in earlier studies~\cite{Csikszentmihalyi2013}. A common characteristic of these studies is the need to deploy unconventional technology, such as beepers and diaries, and train the participants to use them. Furthermore, itself context-oblivious, the technology did not allow the recognition of and sampling at ``interesting" moments only. Finally, relying on the honesty of users' self-reporting is one of the major drawbacks of the traditional ESM approach~\cite{Csikszentmihalyi1987}.

In the past decade, mobile computing devices, including smartphones and wearables such as smart watches, have become a part of everyday life. Integration with the user and high computing and sensing capabilities render these devices a revolutionary platform for social science research~\cite{Miller2012}, where mobile computing can be used to gather fine-grain personal data from a large number of individuals. The availability of these personal data has contributed to the emergence of the new research field of computational social science~\cite{lazer2009life}. 

In this survey we present an overview of user behaviour sampling via mobile computing devices, and pay particularly attention to practical issues associated with designing and deploying behavioural studies using mobile ESM (mESM)\index{Mobile Experience Sampling Method (mESM)}\index{mESM}. First, we identify the main novelties that mESM brings to the table, of which remote sensing is the one poised to induce a major change to the current practice. Then, we survey the most popular open-source tools that streamline study design and deployment, by lifting the burden of mobile device programming from researchers and therapists, who might have a limited set of technological skills. We then discuss the challenges in running an mESM study, including recruiting participants, ensuring non-biased experience sampling, retaining participants, and handling technological limitations of mobile devices. Finally, we present our vision of mESM, which includes adaptive sampling according to a user's lifestyle, delivery of tailored behaviour change interventions based on the sampled data, and proactive reasoning and interaction in the manner of anticipatory computing.

\section{Mobile Computing for ESM}
\label{sec:mobile_esm}

%
\begin{figure}[t]
\sidecaption
\includegraphics[scale=.4]{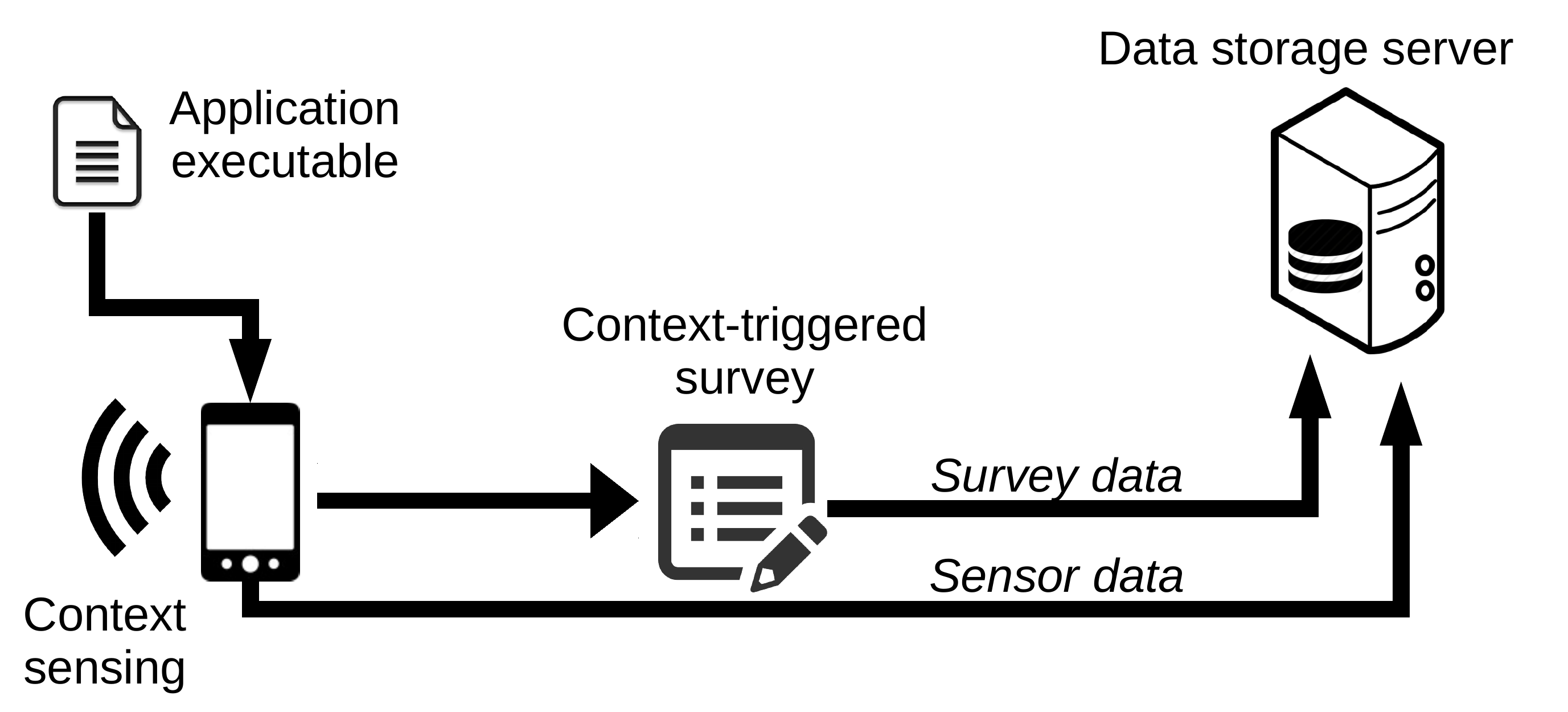}
%
%
\caption{Experience sampling on a smartphone. An ESM application executable is downloaded on the phone. The application manages sensing, and processes the sensed data in order to infer interesting moments when user's data should be captured. When such moments are recognised, a user is prompted to fill in a survey. Data collected from the user, along with the data sensed by mobile sensors, are uploaded to a data storage server for further analysis.}
\label{fig:esm_mobile}       
\end{figure} 
\index{Mobile experience sampling method (mESM)}
Mobile devices are poised to completely transform numerous aspects of experience sampling in behavioural psychology. Study design, participant recruitment, data collection, and the sheer amount of data gathered by mESM are incomparable to the same aspects encountered with the legacy means of experience sampling. As an illustrative example, in Figure~\ref{fig:esm_mobile} we show a smartphone-based mESM application. The application is distributed as an executable file, possibly via an application store, to a large number of participants owning commodity smartphones. A personalised instance of the application is then run at each of the phones, where it harnesses phone's sensing ability to recognise the situation in which a user is, and should the situation be of interest, signals a user to fill in a survey. The user-provided information is then, together with the data sensed by smartphone's sensors, dispatched to a centralised server where it can be analysed.  

Smartphones-based mESM studies improve the traditional beeper and paper form studies in a few important ways. First, unlike beepers and diaries, smartphones are already a part of users' lives, and do not interfere with users' lifestyle. With mESM studies we are ``piggybacking'' in a sense on an already used device, reducing the burden on the participant to carry an additional device, and lowering the cost of the study. Moreover, using a conventional device, participants are less likely to be embarrassed about completing a questionnaire~\cite{Trull2013}. In addition, each ESM study can be carried out as separate mobile application, distributed over a large number of devices via application stores such as Google Play and Apple App Store. This enables unprecedented scalability and parallelisation of experience sampling studies.

Second, modern mobile devices are equipped with a range of sensors, from GPS to light, proximity, and movement sensors. Therefore, unlike beepers, smartphones ``know" the context in which a user is. As shown in the example (Figure~\ref{fig:esm_mobile}), this can augment data collection. ESM studies often aim to capture user experience within a certain situation, for example, whenever a user leaves home. Beepers use preprogrammed times, and in case events cannot be reliably forecast, a user's departure time from a certain location is a likely candidate for such an event, we have no means of ensuring that relevant events are captured. Smartphones, on the other hand, can infer the context from sensor readings, and then prompt the user to fill in the survey as the desired context is happening. Location-dependent reminders, for example, are already a part of commercial applications~\cite{GoogleNow}. In addition, the main caveat of the beeper-based ESM is its reliance on the honesty of self reports. Device location, user's activity, and their social circle, can be inferred with the help of smartphone sensors. Numerous aspects of context can be reported directly by the smartphone, avoiding user-induced errors in the data. Finally, the sensed context can be directly relevant to an ambulatory assessment of a user's psycho-physical state. Previous studies put a great effort to combine participants' diary entries with their heart rate or blood pressure, for example~\cite{Fahrenberg1996}. Nowadays, devices such as smartwatches, which come integrated with with galvanic skin response and heart rate sensors, enable holistic ambulatory assessment/ESM studies at scale.
%

\begin{svgraybox}
\textbf{Mobile Sensing for ESM:} \index{Mobile sensing}Modern smartphones, almost without exception, feature location, orientation, acceleration, and light sensors, together with cameras and a microphone. High-end models host over a dozen of different sensors, including barometer, heart rate and gesture sensors. Combined with high computing power provided by today's phones' multicore CPUs, smartphones represent an attractive platform for real-time context inference. For most of the day phones are carried by their owners, thus sensor data closely reflects actual users' behaviour and the change of the context around the user. With the help of machine learning, personalised models of different contextual aspects can be built on top of the collected data. Phones are routinely exploited to infer users' semantic location (home, work) via GPS-assisted mobility models. Sensor data from a phone's built-in accelerometer can be used to infer a person's physical activity. Sounds captured by the built-in microphone can be processed to infer if a conversation is taking place in a user's vicinity, but also to infer a user's stress level and emotional states~\cite{miluzzo2008sensing,KMMRLA10:EmotionSense,2012stresssense}. A Bluetooth chip, itself merely a short-range communication enabler, can be used to infer social encounters of a phone owner~\cite{Rachuri2011}. A number of high-level descriptors of human behaviour can be inferred by combining the sensor data coming from different sensors, including contextual information from online social networks~\cite{Lane2010, Mehrotra2014}. However, we should not forget that above all, today's phones are communication devices providing always-on voice and data connectivity. Thus, for the first time, with mobile computing ESM researchers have a possibility to design truly context-aware studies, to get real-time information about the context in which the participants are, and to adjust sampling strategies on the fly.
\end{svgraybox}

\section{Modern mESM Software Frameworks}
\label{sec:frameworks}
\index{Frameworks for mobile experience sampling}
The design, implementation and deployment of experience sampling studies via mobile devices requires expertise that is not confined to the traditional social science training. A smartphone-based mESM study, for example, entails a significant programming effort in building the application and managing mobile sensing, as well as the construction of sophisticated machine learning models for context inference, and ensuring reliable data transfer from remote devices to a centralised server. Not only are these tasks often outside the psychological researchers' and therapists' expertise, they also result in a lot of replicated effort for each new study. 
\begin{sidewaystable}
\caption{Frameworks for building mobile experience sampling studies.}
\label{tab:frameworks}       
%
%
\vspace{4in}
\begin{tabular}{p{2cm}p{4.1cm}p{1.5cm}p{1.8cm}p{2cm}p{1.5cm}p{1.5cm}p{1.3cm}p{4cm}}
\hline\noalign{\smallskip}
\textbf{Name} & \textbf{URL} & \textbf{Platform} & \textbf{Surveys\newline support} & \textbf{Context\newline sensing} & \textbf{Mobile component} & \textbf{Server component} & \textbf{Data\newline analysis} & \textbf{Description}  \\
\noalign{\smallskip}\svhline\noalign{\smallskip}
ESP~\cite{Barrett2001} & \url{www.experience-sampling.org} & Palm Pilot & Yes & No & Yes & Yes & No &\begin{tabular}[l]{@{}l@{}}  The first mESM app,\\PC-based study design tool.\end{tabular}\\\hline
MyExperience \cite{Froehlich2007} & \url{myexperience.sourceforge.net} & Pocket PC & Yes & Yes & Yes & No & No & \begin{tabular}[l]{@{}l@{}}  Introduces context-aware\\sampling.\end{tabular}\\\hline
PsychLog~\cite{Gaggioli2013} & \url{sourceforge.net/projects/psychlog} & Windows mobile & Yes & Yes & Yes & No & No & \begin{tabular}[l]{@{}l@{}}  Supports external\\ sensors (e.g. ECG).\end{tabular}\\\hline
AndWellness \cite{Hicks2010} & Not available (July 2015) & Android & Yes & Yes & Yes & Yes & Yes & \begin{tabular}[l]{@{}l@{}}  Health care-oriented,\\data visualisation.\end{tabular}\\\hline
EmotionSense \cite{SensorManager,Lathia2013b} & \url{emotionsense.org}  & Android & Yes\newline(unreleased) & Yes & Yes & Yes\newline(unreleased) & No & \begin{tabular}[l]{@{}l@{}}  Emotion sensing app,\\with open-source libraries.\end{tabular}\\\hline
Ohmage~\cite{Ramanathan2012} & \url{ohmage.org} & Android/\newline iOS & Yes & Yes & Yes & Yes & Yes & \begin{tabular}[l]{@{}l@{}} Allows high-level\\context inference.\end{tabular}\\\hline
funf~\cite{Aharony2011} & \url{www.funf.org} & Android & No & Yes & Yes & No & No & \begin{tabular}[l]{@{}l@{}}A rich sensing\\ framework.\end{tabular}\\\hline
OpenDataKit \cite{Hartung2010} & \url{www.opendatakit.org} & Android & Yes & Yes & Yes & Yes & Yes & \begin{tabular}[l]{@{}l@{}}Data collection tool\\targeting non-expert users.\end{tabular}\\\hline
Paco & \url{github.com/google/paco} & Android/\newline iOS & Yes & Yes & Yes & Yes & No & \begin{tabular}[l]{@{}l@{}}  An extensible framework for\\ quantified self experiments.\end{tabular}\\\hline
Purple Robot & \url{tech.cbits.northwestern.edu/\newline purple-robot} & Android & No & Yes & Yes & Yes & No & \begin{tabular}[l]{@{}l@{}}  A framework for sensing\\and sensor-based actioning. \end{tabular}\\\hline
SenSocial~\cite{Mehrotra2014} & \url{cs.bham.ac.uk/~axm514/sensocial} & Android & No & Yes & Yes & Yes & No & \begin{tabular}[l]{@{}l@{}}  A library for joint sampling of\\ OSN and sensor data streams. \end{tabular}\\
\noalign{\smallskip}\hline\noalign{\smallskip}
\end{tabular}
\end{sidewaystable}

Table~\ref{tab:frameworks} lists some of the frameworks developed by the research community in order to streamline the process of conducting mobile experience sampling studies\footnote{Every effort has been made to provide truthful descriptions of the listed mESM frameworks, however, due to limited documentation and publications related to some of the frameworks the listed properties should be taken with caution.}\footnote{The goal of this article is to suggest guidelines for future research in the field, thus we concentrate on free open-source software developed in academia, as such software can serve as a basis for next generation mESM frameworks. Commercial products for supporting mESM are outside of the scope of our article.}. The first frameworks preceded the smartphone era. \textit{The Experience Sampling Program (ESP)} runs on Palm Pilot PDA devices, and lacks the sophisticated context awareness introduced in later frameworks~\cite{Barrett2001}. However, the ESP was the first framework to introduce an authoring tool for designing experience sampling questionnaires for mobile devices. The tool also lets a study designer define a logic for timing the questionnaire prompts. Compared to the traditional beeper and diary studies, ESP-based studies combine signalling and data collection on the same device, yet, PDA devices have never achieved mass popularity needed for large-scale ESM studies in the wild.

Recognising context was the most important missing feature in traditional ESM studies. Event-contingent sampling, where the time of sampling depends on the context or an event in which a user is, is of particular interest for psychological studies~\cite{Reis2000}. Such sampling is important in case target events are rare, short-lasting, or unpredictable, in which case periodic sampling might completely miss them. For example, Cote and Moskowitz investigated the impact of the ``big five" personality traits on the relationship between interpersonal behaviour and affect~\cite{Cote1998}. The participants were instructed to fill out a questionnaire following each interpersonal interaction. Without context-aware devices, Cote and Moskowitz used beeper to periodically remind participants to keep up with the study, but telling them that the answers should be provided only after an interaction has happened. However, the correctness of this approach, particularly the timeliness of harvested data, depends on the users' compliance with the rules of the study. The study designers have no means of checking whether, and when, interpersonal interactions have happened.

The \textit{MyExperience} framework~\cite{Froehlich2007}, built upon an earlier context-aware mESM tool developed by Intille et al.~\cite{Intille2003}, runs on Pocket PC and lets researchers design studies that embrace context awareness provided by mobile sensing. On one side, sensor data can be passively logged on the user device and uploaded to a server, on the other, the data can be processed on the phone to infer the context and trigger event-contingent sampling if needed. As one of the first examples of an mESM framework, MyExperience's context triggering relies on raw sensor data, i.e., it does not perform any inference in order to extract higher level information. For example, the framework supports sampling when a user moves from one mobile cellular base station to another, but cannot recognise if a user arrived at a semantically significant location, say his/her workplace.

MyExperience and ESP set a foundation for modern experience sampling frameworks, while a wider adoption of mESM frameworks came with the rise of the smartphone that enabled remote data gathering without requiring user actions, and context-aware user querying. The first Apple iPhone, released in 2007 and packed with numerous high-resolution sensors, marked a revolution in mobile sensing. Devices from different vendors followed, often running Android OS that enabled finer control over sensing than ever before. Modern smartphone sensing, however, has to balance between limited energy resources available on the phone, and the need for fine-grained data from multiple sensors. In addition, smartphone's sensors were not originally conceived for continuous sampling. Sensing and data collection management become a new pressing issue for mobile computing. Sensing frameworks such as \textit{ESSensorManager} (a part of the Emotion Sense project) aim to abstract the details about data acquisition and collection from an application developer and automate sensing as per predefined policies~\cite{SensorManager}. The \textit{funf} framework adds an option of basic survey data collection, and was used in a detailed 15-month long study of 130 participants' social and physical behaviour~\cite{Aharony2011}. The study provided an in-depth investigation of the connection between individuals' social behaviour and their financial status, and the effects of one's network in decision making. The power of sensed data was further demonstrated when on top of funf the authors built an intervention that not only sampled users' behaviour, but also influenced users to exercise more. 

Raw data from mobile sensors can be difficult to interpret in terms that are of direct interest when sampling human behaviour. High-level inferences often need to be made before sampled experience becomes valuable for researchers and therapists. \textit{Purple Robot}'s authors claim that their framework supports statistical summaries of the user's communication patterns, including phone logs and text-message transcripts. \textit{Ohmage}~\cite{Ramanathan2012}, a platform for participatory sensing and ESM studies, hosts a few data classifiers that can infer concepts such as mobility and speech. A psychology practitioner faces a large barrier between raw data from sensors that are related to users' behaviour, e.g., their movement, and the high-level labels of those behaviours, e.g., if people are walking or driving a car. It is crucial for mESM frameworks to abstract the sensing, in this case accelerometer and GPS sensor sampling, data processing, in this case extracting acceleration variance and GPS-reported speed, and machine learning, in this case classify a mobility mode, and deliver high-level information. Recently, Google released its activity recognition API for Android, enabling easy, albeit crude, inference of user's activity state~\cite{GAR}.

Besides built-in smartphone sensors, \textit{PsychLog}~\cite{Gaggioli2013} and \textit{Open Data Kit (ODK)}~\cite{Hartung2010} frameworks provide support for external sensors that can be attached to a mobile devices, greatly enhancing the utility of ESM for ambulatory assessment. Moreover, for understanding human behaviour, any source of human-related information can be a sensor. In particular, social interactions are for a large part conducted over online social networks (OSNs), and monitoring OSNs is increasingly becoming a focus of social science studies. For instance, this information-rich sensor can be merged with the physical context sensors, in order to uncover socio-environmental relationships. An example of system supporting this type of real-time data fusion is \textit{SenSocial}, a distributed (residing on mobiles and a centralised server) middleware for merging OSN-generated and physical sensor data streams~\cite{Mehrotra2014}. 

A successful mESM framework abstracts mobile application programming from an intervention developer, yet exposes enough functionality so that a variety of studies are supported. Early on, MyExperience used an XML-based interface through which study developers can define sensor data to be collected, survey questions and triggers that will alert users to fill in the surveys. Although close to a natural language, XML scripts are not an ideal means of describing a potentially complex mESM application. Targeting primarily less tech savvy users in the developing world, ODK puts an emphasis on hassle-free study design process~\cite{Hartung2010}. The framework introduces a survey and sensor data collection authoring tool that enables drag-and-drop study design. ODK is further tailored for non-expert designers and study participants by supporting automated data upload, storage and cloud transfer, as well as automated phone prompts that users respond to with keypad presses. The Project Authoring tool that comes with the Ohmage framework guides a study designer through a project definition process, and outputs an XML definition of the study. Ohmage also features tools such as Explore Data, Interactive and Passive Dashboards, and Lifestreams, that enable in-depth analysis and visualisation of collected data. In particular, Lifestreams use statistical inference on raw collected data in order to examine behavioural trends and answer questions such as ``how much time a user spends at work/home?".

\section{Challenges of Sampling with Mobile Devices}
\label{sec:challenges}

Contextual data collection, rapid prototype design, study scalability, and automated result analysis are just some of the ways in which mobile devices revolutionise the traditional ESM. However, certain original limitations of the method are still present even with this new technology. How to capture experience without interfering with the participant's lifestyle, and how to sample relevant moments when the user's lifestyle is highly varying and unpredictable are some of the questions existing since the ESM was introduced. Some other issues, such as user recruitment become more prevalent now that a study can be distributed in form of an application that could be run on millions of devices. Furthermore, the new platform introduced novel technical challenges that impact the way studies should be designed. 

\subsection{Recruiting and Maintaining Participation}
\index{Recruitment of participants}
The Internet empowered social scientists with an easy access to a large and diverse pool of participants, alleviating the predominant issue of running psychological studies on a small group of college students~\cite{Gosling2004}. Yet, recruitment in initial mESM studies saw little benefit from the Internet, since the participation was throttled, just like in the case of the older beeper technology, by the availability of the supporting hardware, i.e., Pocket PCs. Nowadays, with 1.5 billion users the smartphone is one of the most ubiquitous devices on the planet. Consequently, smartphone-based mESM studies can be distributed at an unprecedented scale. 

All major smartphone operating systems, such as Android, iOS and Windows Mobile, have their corresponding online application stores. With these stores as distribution channels, the pool of study participants is no more confined to a certain population that a study designer can reach. Despite concerns about the diversity of participants recruited through the Internet, Gosling et al. show that such a sample is more representative of the actual demographics than a sample recruited through traditional means~\cite{Gosling2004}. Note, however, that the Internet has been around longer than smartphones, and has penetrated almost all segments of the society. Still, a rapid rise in smartphone ownership promises to erase any demographic biases that currently may exist in smartphone usage. 

To a potential participant, a smartphone-based mESM provides an additional benefit of anonymity, as a user does not have to disclose her real name, nor needs to meet the people/organisation running the study. On the down side, researchers have to sacrifice the close control over who the study participants are. For example, there is no reliable way to confirm that a person's age is truthfully reported, potentially allowing minors to run adult-only studies. Mobile sensing can somewhat ameliorate the problem of false reporting, as it provides information about users' activity, movement, geographic location, communication patterns and others. It has been shown that such information reflects users' age, gender, social status~\cite{Martinez2012}. Besides assisting in the pruning of false reports, the link between sensor data and the demographics can be used to selectively target a certain demographic group, or to tailor the study according to different groups, e.g., adjusting sampling times according to local customs, sending different questions to people belonging to different social groups.

A low entry barrier that smartphone mESM applications provide, also means that leaving a study is easy -- a user just has to uninstall or ignore the application. Usually only a percentage of users is active after the first initial period. 
Furthermore, on global application markets mESM study applications compete with hundreds of thousands of useful and fun applications. One way of attracting and retaining a wide audience for an mESM study is by providing some kind of information back to the user. In Emotion Sense, an mESM application that captures emotional state and contextual sensor information, participant retention is achieved through gamification and provision of self-reflecting information about the user~\cite{Lathia2013b}. Emotion Sense invites a user to ``unlock" different parts the application by providing further experience samples.

Another means of attracting and retaining users is through remuneration. Amazon Mechanical Turk is a popular crowdsourcing marketplace where \textit{requesters} post jobs to be completed by \textit{workers}. The jobs typically consists of simple, well-defined tasks for which computers are not suitable, such as data verification, image analysis and data collection. Workers are paid a previously agreed sum of money per completed task. Any adult person, from any continent, can become a worker. In~\cite{Mason2012}, Manson and Suri discuss the opportunities for conducting behavioural research on Amazon's Mechanical Turk. A wide pool of participants for a study and a payment system that enforces job completion are emphasised as the main advantages of the Mechanical Turk. On the other hand, artificial automatic workers -- bots -- can be used by workers to fake study results without actually running the study. In addition, the Mechanical Turk workers are not representative of general, even online, demographics. We believe, however, that mobile sensing can be used to ameliorate both problems. Artificial behaviour can be detected through unusual activity and movement patterns, while as explained earlier, sensed data can be used to infer the participants demographics.

\subsection{Sampling at the Right Time}
\index{Interruptibility}
Smartphone-based mESM applications run on devices that are an inseparable part of participants' lives. Thus, it is crucial for a sampling schedule to be in harmony with the user's lifestyle. A well designed interruption schedule can help in both retaining users, but also in fulfilling the true role of experience sampling -- recording momentary experience -- as participants not wishing to be interrupted are likely to introduce the recall bias by delaying their answers until they find a suitable moment to respond~\cite{Mehrotra2015}.

Interactivity is in the core of human behaviour, as we balance between working on a task and switching to other pressing issues. The mobile phone makes our lives increasingly interactive as notifications delivered via mobile devices became a dominant means of signalling possible tasks switching events. In mESM studies mobile notifications are used to prompt users to fill in sampling surveys. The timing of notifications is important, since in case a notification arrives in an opportune moment for interruption the user reacts to it quickly, and fills in a survey with timely data. Several research studies investigated mobile notification scheduling in order to identify these opportune moments, and found that the context in which a person is, to a large extent, determines if a user is interruptible or not~\cite{Ho2005,Hofte2007}. Equipped with sensors, a mobile device can infer this context. Ho and Intille, for example, show that external on-body accelerometers can detect moments of activity transitions, and that in such moments users react to an interruption more favourably~\cite{Ho2005}. 

In~\cite{Pejovic2014} a 20-person two-week study of mobile interruptibility shows how data from built-in smartphone sensors relates to user's attentiveness to mobile interactions in form of notifications. The study demonstrates that a personalised model of the sensed data -- interruptibility relationship can be built, after which the authors extract the sensor modalities that describe user interruptibility, including acceleration, location and time information, and implement personalised machine learning models that, depending on the given sensor input, infer interruptibility. The findings are funnelled into a practical system termed \textit{InterruptMe} -- an Android library for notification management, that informs an overlying application about opportune moments in which to interrupt a user\footnote{InterruptMe is available as a free open-source software at \url{bitbucket.org/veljkop/intelligenttrigger}}.  Machine learning-based models that InterruptMe builds are refined over time. However, ESM studies are limited by the number of samples that are taken from a single person over a period of time, thus the phone should learn about when to interrupt a user with as few training samples as possible. Kapoor and Horvitz propose a decision-theoretic approach for minimising the number of samples one needs to take from a user in order to build a reliable model of that persons interruptibility~\cite{Kapoor2008}. Finally, the InterruptMe study finds that interruption moments cannot be considered in isolation, and that users' sentiment towards an interruption depends on the recently experienced interruption load. This may become a limiting factor as the number of applications, and consequently notifications, that a user gets on her phone grows.

\subsection{ESM Studies and Contextual Bias}
\index{Contextual bias in experience sampling}
A decision about when, or under which conditions, a notification to complete a survey should be fired is not only important for improved user interaction and compliance, it also has a crucial effect on the data that will be harvested. 

Studies have, for example, been designed to collect data at random intervals \cite{Barrett2001} or when smartphone sensors acquire readings of a particular value\cite{Froehlich2007}. While the latter is often motivated by directly tying a device state with a device-related assessment (e.g., plugging in the phone triggering questions about phone charging \cite{Froehlich2007}), both of these methods have been used by researchers to make inferences and test hypotheses about broad, non-device specific aspects of participants' behaviours, such as daily events and moods \cite{Clark1988} and sustainable transportation choices \cite{Froehlich2009}.

This methodology assumes that the design choice of which trigger to use will not affect (or, indeed, will even augment the accuracy of) the contextual data that can be used to learn about participants. In doing so, these studies do not take into account the effect that the designed sampling strategy has on the conclusions they infer about participants' behaviours. However, these behaviours are likely to be habitual or, more broadly, variantly distributed across each day. For example, since people may split the majority of their time between home and work, sampling randomly is likely to fail capturing participants in other locations. While this could easily be solved by using survey triggering that is conditioned on the value of a user's location (from here on termed \textit{location-based triggering}), it is not clear how doing so affects sampling from the broader set of sensors that researchers may be collecting data from (i.e., how would location-based sampling bias the data about participants activity levels?). 

In \cite{Lathia2013b}, Lathia et al. study the effect of mESM design choices on the inferences that can be made from participants' sensor data, and on the variance in survey responses that can be collected from them. In particular the authors examined the question: {\em are the behavioural inferences that a researcher makes with a time or trigger-defined subsample of sensor data biased by the sampling strategy's design?} The study demonstrates that different single-sensor sampling strategies will result in what is refer to as \emph{contextual dissonance}: a disagreement in how much different behaviours are represented in the aggregated sensor data.

To analyse this, Lathia et al. examine the extent that studies' design influences the response and sensor/behavioural data that researchers can collect from participants in context-aware mESM studies. If the design of the experiments does not have any influence on the data that are collected, we would expect that, on aggregate, the data gleaned from different designs would be consistent with one another. Instead, the study demonstrates that different single-sensor sampling strategies result in contextual dissonance, i.e., a disagreement in how much different behaviours are represented in the aggregated sensor data. This conclusion is based on a 1-month, 22-participant mESM study that solicited survey responses about participants' moods while collecting data from a set of sensors about their behaviour. This falls under two broad groups:
\begin{itemize}
	\item \textit{Amount of Data.} Using different sensors to trigger notifications will directly impact the amount of data that researchers can collect. In this study, microphone-based triggers, which pop-up a survey only if a non-silent audio sample is sensed, produce a higher per-user average number of notifications; conversely, the communication-based triggers, fired after a call/SMS received/sent event, produce the lowest average number of notifications per participant ($5.11 \pm 3.69$), indicating that participants were not using their phone for its call/SMS functionality throughout the day.
	\item \textit{Response Data.} In addition, the study examines how the feelings of participants varied under different experimental conditions. In this case, the null hypothesis is that the \emph{design} of survey triggers does not bias the resulting sample of affect data that is collected. This hypothesis is rejected, with varying levels of confidence, if the resulting p-values are small. In 4 of the 6 tests that were performed, it was found that the negative affect ratings (and 2 of 6 for the positive ratings) were significantly different from one another with at least 90\% confidence. Uncovering why this result has emerged calls for further research: it may be explained by the fact that some triggers were likely to be more obtrusive than others, thus affecting responses. 
\end{itemize}

Finally, we point out that the above conclusion is based on a time-limited and small-scale study. Perhaps some of these challenges can be overcome by using larger populations for longer times. However, these results stand as a warning for researchers to be mindful of in their future mobile experience sampling studies.

\subsection{Technical Challenges in mESM}

A smartphone is a multi-purpose device used for voice, video, and text communication, Web surfing, calendar management or navigation, among other applications. This versatility puts pressure on smartphone's resources, and limits its usability for experience sampling. Furthermore, unlike a conventional mobile application, an mESM application needs to be always-on, sense the context and sample users' experiences as necessary. 

One of the key constraints of many mobile sensing applications is a limited capacity of a mobile device's battery. Power-hungry sensors, such as a GPS chip, were not envisioned for frequent sampling. Adaptive sensing is a popular means of reducing energy requirements of a mobile sensing application. Here, samples are taken less frequently, or with a coarser granularity, e.g., a user's location is recorded with a base station ID, instead of with accurate GPS coordinates. \textit{AndWellness} framework, for example, lets study designers tune the balance between the sampling resolution and power drain~\cite{Hicks2010}. The same framework implements hierarchical sensor activation, another means of minimising energy usage. This approach uses low-power, yet less accurate, sensors in order to infer if high-power, fine-grain sensors should be turned on. An example of hierarchical sensing can be seen in AndWellness: a change in a device's WiFi access point association serves as an indicator of user's movement, and if movement is detected a GPS chip is activated. While Ohmage implements adaptive sensing to save energy during speech detection by adjusting the sampling rate depending on the sensing results -- if the app has not detected speech over a certain amount of time, it exponentially decreases the sampling rate. How to adjust the sampling rate with adaptive sensing, or hierarchical activation, while ensuring that the events of interest are not missed is an open research question.  In SociableSense, a mobile application that senses socialisation among users \cite{Rachuri2011}, a linear reward-inaction function is associated with the sensing cycle, and the sampling rate is reduced during ``quiet" times, when no interesting events are observed. The approach is very efficient with human interaction inference, since the target events, such as conversations, are not sudden and short; for other types of events, different approaches might be more appropriate.

Client-server architecture is at the core of almost every ESM framework. Servers are used for centralised data storage and analysis, data visualisation, and remote configuration of sampling mobiles. The balance of functionalities over mobiles and a server can have a significant impact on the performance and the possibilities of an mESM application. For automated labelling of human behaviour, say recognising if a person is walking or not, a substantial amount of data, in this case accelerometer data, has to be processed through machine learning models. Server-based processing comes with benefits of a high computational power of multicore CPUs, and a global view of the system, and as a consequence data from all the users can be harnessed for individual (and group) inferences. On the other hand, the transfer of the high-volume data produced by mobile sensors can be costly, especially if done via a cellular network. Some mESM and mobile sensing frameworks, including Ohmage and ESSensorManager, let developers define data transfer policies, such as ``upload mobile data only via Wi-Fi" and ``do not upload any data if the battery level is below 20\%".

Besides its impact on performance and cost, balancing local and remote processing is important for ESM studies due to privacy issues associated with data transfer and storage (see for example~\cite{de2014openpds}). Location, video and audio data are particularly vulnerable, yet can be protected with a suitable balance of remote and local processing. For example, if we want to infer that a user is having a conversation, instead of transferring raw audio data for server-side processing, we can extract sound features relevant for speech classification, such as Mel-Frequency Cepstral Coefficients (MFCC) of sound frames that contain sounds over a certain threshold intensity, and send these for remote analysis. Even if a malicious party gets access to this data, the original audio recording cannot be reconstructed. Similarly, instead of sending raw geographic coordinates to a server, a mobile application could host an internal classifier of a user's semantic location (e.g., home/work), and send already processed data, minimising the amount of information about the user that can be revealed.

\section{Future of mESM}
\label{sec:future}

Mobile computing is rapidly transforming social sciences. First, the range of potential study participants has expanded dramatically. Nowadays researchers have access to a virtually world-wide pool of participants. Second, the granularity of personal data gathered through mobile sensors and phone-based interactions, including online social network activities, can paint a very detailed picture of an individual's behaviour. In addition, long-term data can be obtained, as long as the mESM application manages to keep users engaged, and they do not remove the application from their phones. 

The above transformation requires the rethinking of the traditional social science approaches. Computational social science~\cite{lazer2009life} has emerged as a field that harnesses statistical and machine learning approaches over user-generated ``big data" in order to explain social science concepts, including behaviour. This field is inherently interdisciplinary and rather broad, since it involves computer scientists, engineers, and social scientists, who traditionally had limited interactions in the past.  

\subsection{Integration with Behaviour Interventions}

Ubiquitous mobile computing devices and the ESM provide a detailed assessment of human behaviour at an unprecedented scale. A natural next step is to use the information about the existing individual and group behaviour to affect future behaviour. Behaviour change interventions (BCIs) are a psychological method that aims to elicit a positive behaviour change. These interventions commonly include collecting relevant information about the participant, setting goals and plans, monitoring behaviour, and providing feedback. Digital BCIs (dBCIs)\index{Digital behaviour change interventions} moved behaviour change interventions to the Web. The benefits of this transition include increased control over the content and the time of information delivery of information, as well as the reduction in the intervention cost, since the need to a face-to-face interaction with a therapist is avoided. In addition, dBCIs open an opportunity for scalable automatic content tailoring. Such tailoring has been shown to be effective for the actual behaviour change~\cite{Yardley2010}. 

\begin{figure}[h]
\centering
\includegraphics[width=0.85\textwidth]{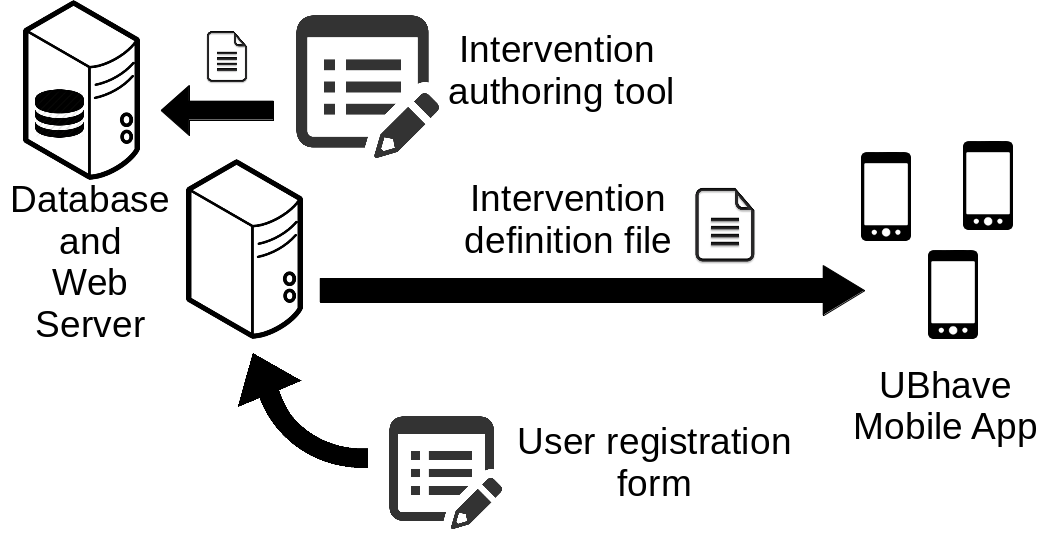}
\label{fig:plan}
\caption{Overview of the UBhave framework for mobile digital behaviour change interventions.}
\label{fig:UBhave_overview}
\end{figure}

Recently, both isolated~\cite{Morrison2014} and systematic~\cite{Lathia2013} attempts have been made to move dBCIs from the Web to smartphones. The new platform enables intervention content delivery anywhere and anytime. In addition, a personalised use of the phone indicates that through mobile sensing and user sampling a detailed personalised model of user behaviour can be constructed and used to drive the intervention. For example, users whose samples indicate sedentary behaviour can be provided with a positive feedback whenever they are detected to be active. Technical difficulties in building an integrated mESM and intervention distribution method hamper proliferation of mobile dBCIs. The system design and programming effort associated with implementing a system for remote mobile sensing and experience sampling, information delivery, user management and personalised behaviour modelling is overwhelming. Certain existing frameworks, such as AndWellness~\cite{Hicks2010} and BeWell~\cite{Lane2011bewell}, concentrate on sampling data relevant for users' health and well-being, yet none of them cater specifically to dBCIs, and none of them solves the above technical difficulties. The UBhave framework (Figure~\ref{fig:UBhave_overview}) aims to overcome this by providing out-of-the-box support for mobile dBCI design and deployment~\cite{Hargood2014}. The framework consists of an intervention authoring tool, through which therapists can design interventions, and an automated translation tool, which translates the design into an intervention file.
This file is then deployed to and interpreted by participating mobile devices running the UBhave client application. The framework ensures that therapist's instructions on when to sample user experience and sensor data are followed by the mobile app, and that the behaviour changing advice is delivered to the user when needed. 

The UBhave framework\index{UBhave framework}  is the first that extends the idea of mESM beyond behaviour tracking to behaviour change. While the idea of mobile dBCIs sounds extremely promising, only after a wider adoption and broader behaviour change studies it is will be possible to quantify the actual effectiveness of mobile interventions.

\subsection{Anticipatory mESM}
\index{Anticipatory mESM} 
The awareness of the current context is the main novelty of experience sampling using mobile devices. A prediction of future context has a tremendous potential to make an mESM a key tool for explaining human behaviour. Although predicting, and even inferring, participants' internal states with mobile sensors is yet to be achieved, prediction of some other behavioural aspects, such as users' movement trajectories or calling patterns, has already been demonstrated~\cite{Phithakkitnukoon2011,Sadilek2012}.

Anticipatory computing\index{Anticipatory computing}  systems rely on the past, present and predicted future information to bring judicious decisions about their current actions~\cite{Rosen1985}. Mobile ESM applications could, in an anticipatory computing system manner, intelligently adapt their sampling schedules based on the predicted user behaviour. For example, an mESM application that could anticipate a depressive episode, could adapt its sampling to capture, with fine resolution, behaviour and the context just before the event of interest. This would not only provide very detailed information about the context that lead to the onset of a depressive episode, but also use phone's battery resources more efficiently. 

Finally, we also envision proactive digital behaviour interventions delivered via mobile devices~\cite{Pejovic14:anticipatory}. Besides the sampling schedule, anticipatory dBCIs would also adapt the feedback they give to a user according to the predicted state of the user, and the predicted effect the feedback will have on the user. For example, a smart wristband occasionally samples a user's heart rate. Based on the readings, the system, encompassing a phone and a wristband, predicts that the user is in risk of being highly stressed out. The system accesses user's online calendar and examines tasks scheduled for today. Then, it intelligently reschedules tasks to alleviate the risk of high stress and suggests a new schedule to the user. Technical obstacles associated with this scenario include stress prediction, itself quite challenging, but also the prediction of how a user will react to a given change in the calendar. Will the change really help alleviate stress? The idea of anticipatory mobile computing has just recently appeared in the literature, while the conventional mobile dBCIs have not yet taken off. Therefore, we are yet to witness anticipatory mobile dBCIs in practice.

\section{Conclusion}
\label{sec:conclusions}

Smarthphones and other mobile devices, such as wearables, are the first sensing and computing devices tightly interwoven into our daily lives. They represent revolutionary platforms for social science research and for the emerging field of computational social science. They indeed open a window of opportunity for social scientists to learn about human behaviour at previously unimaginable granularity and scale. A wide span of behaviours can be captured via mobile experience sampling. This covers the domains for which the traditional experience sampling has already been employed, such as studies of a personal time usage~\cite{Kubey1990}, emotions of different stigmatised groups~\cite{Frable1998}, and classroom activities~\cite{Turner1998}, to name a few. In addition, mobile computing enables the investigation of new domains such as the location-dependent privacy management of sharing information on online social networks~\cite{Abdesslem2010}, sleep monitoring~\cite{Lane2011bewell} and mobile application evaluation~\cite{Consolovo2003}. Furthermore, mESMs are being used for ambulatory assessment in areas that span from sexual behaviour to physical exercise monitoring\footnote{The following URL lists currently running experience sampling projects using the Ohmage framework: \url{http://ohmage.org/projects.html}}.

We highlighted key benefits of mobile experience sampling, and presented mESM frameworks that abstract the technical effort of building a sensing and sampling mobile application, and enable seamless implementation and deployment of large scale social studies. Our vision for the evolution of mESM goes along the lines of the general consensus that mobile applications need to be ``stealth", perfectly integrated with everyday lives. Therefore, we envision considerate mESM studies, where the interaction with the user is minimally invasive, and aligned with the sensed user behaviour. Furthermore, harnessing the persuasive power of the smartphone, we also see proactive behaviour change interventions based on the automated analysis of the sampling results. Finally, outside of the scope of this review, but important for the ecosystem of users, intervention designers and mESM framework developers are the questions of large-scale data mining, interpretation and visualisation, data feedback to study participants, and privacy and ethics issues associated with mobile sensing.

\begin{acknowledgement}
This work was supported through the EPSRC grants ``UBhave: ubiquitous and social computing for positive behaviour change" (EP/I032673/1) and ``Trajectories of Depression: Investigating the Correlation between Human Mobility Patterns and Mental Health Problems by means of Smartphones" (P/L006340/1).
\end{acknowledgement}

\bibliographystyle{abbrv}
\bibliography{references}
\end{document}